\documentclass[iop]{emulateapj}
\usepackage{amsmath}
\usepackage{apjfonts}
\usepackage{color}
\usepackage{verbatim}
\usepackage{hyperref}
\usepackage[utf8]{inputenc}
\newcommand{\dof}{{\rm dof}}

\shorttitle{Modeling Hot Gas Flow in NGC3115}
\shortauthors{Shcherbakov, Wong, Irwin, \& Reynolds}

\begin{document}
\title{MODELING HOT GAS FLOW \\
IN THE LOW-LUMINOSITY ACTIVE GALACTIC NUCLEUS OF NGC3115}

\author{Roman V. Shcherbakov\altaffilmark{1,2}, Ka-Wah Wong\altaffilmark{3}, Jimmy A. Irwin\altaffilmark{3}, Christopher S. Reynolds\altaffilmark{1}}

\altaffiltext{1}{Department of Astronomy, University of Maryland, College Park, MD 20742-2421, USA}
\altaffiltext{2}{Hubble Fellow}
\altaffiltext{3}{Department of Physics and Astronomy, University of Alabama, Box 870324, Tuscaloosa, AL 35487, USA}

\begin{abstract}
Based on the dynamical black hole (BH) mass estimates, NGC3115 hosts the closest billion solar mass BH.
Deep studies of the center revealed a very underluminous active galactic nucleus (AGN) immersed in an old massive nuclear star cluster.
Recent $1$~Ms \textit{Chandra} X-ray visionary project observations of the NGC3115 nucleus resolved hot tenuous gas, which fuels the AGN.
In this paper we connect the processes in the nuclear star cluster with the feeding of the supermassive BH.
We model the hot gas flow sustained by the injection of matter and energy from the stars and supernova explosions.
We incorporate electron heat conduction as the small-scale feedback mechanism, the gravitational pull of the stellar mass, cooling, and Coulomb collisions.
Fitting simulated X-ray emission to the spatially and spectrally resolved observed data, we find the best-fitting solutions with $\chi^2/\dof=1.00$ for $\dof=236$ both
with and without conduction. The radial modeling favors a low BH mass $<1.3\times10^{9}M_\odot$. The best-fitting supernova rate and the best-fitting mass injection rate
are consistent with their expected values.  The stagnation point is at $r_{\rm st}\lesssim1$~arcsec, so that most of gas, including the gas at a Bondi radius $r_B=2-4$~arcsec,
outflows from the region. We put an upper limit on the accretion rate at $2\times10^{-3}M_\odot{\rm yr}^{-1}$.
We find a shallow density profile $n\propto r^{-\beta}$ with $\beta\approx1$ over a large dynamic range.
This density profile is determined in the feeding region $0.5-10$~arcsec as an interplay of four processes and effects:
(1) the radius-dependent mass injection, (2) the effect of the galactic gravitational potential, (3) the accretion flow onset at $r\lesssim1$~arcsec, and (4) the outflow at $r\gtrsim1$~arcsec.
The gas temperature is close to the virial temperature $T_v$ at any radius.
\end{abstract}


\section{INTRODUCTION}\label{sec:introduction}
Both theory and observations indicate that a typical active galactic nucleus (AGN) is not particularly active \citep{Ho:2008rev}.
A median Eddington ratio of $\lambda=L_{\rm bol}/L_{\rm Edd}\sim10^{-5}$ is found in a distance-limited Palomar survey of the nearby AGNs \citep{Ho:2009se},
so that most galactic nuclei are inactive at any given time and any given nucleus is inactive most of the time. The observed short AGN duty cycle \citep{Greene:2007aa}
is readily explained by the large-scale feedback shutting off the central engine soon after an active phase begins \citep{Hopkins:2009kq} leading to the so-called low-luminosity (LL) AGN.

The theory of the gas flow in LLAGNs has been studied for over $60$ years. In their seminal work \citet{Bondi1952} introduced
a characteristic radius of the black hole (BH) gravitational influence now called the Bondi radius $r_B=2G M_{BH}/c_s^2$,  where $c_s$ is the adiabatic sound speed.
Since then BH feeding is traditionally associated with processes near the Bondi radius.
It was uncovered over the years that the Bondi model has a limited applicability to LLAGNs.
\citet{Quataert:2000aj} showed that there may exist a smooth transition at $r\sim r_B$ from the galactic flow to the accretion flow governed by a transition from the galactic gravitational potential
to the BH potential. \citet{Shcherbakov:2010cond} showed that the gas starting at the Bondi radius may not settle into an inflow, but instead be a part of an outflow.
Various models were proposed for the inflow such as advection-dominated accretion flows (ADAFs) \citep{Narayan:1995kj},
convection-dominated accretion flows (CDAFs) \citep{Narayan:2000tr,Quataert:2000er}, and adiabatic inflow-outflow solutions (ADIOS) \citep{Blandford:1999}.

LLAGNs are fed via a variety of the mechanisms. First, the gas traveling from galactic scales may form an inflow onto the BH \citep{Hopkins:2006mi}.
Galaxies with a large gas content such as our own spiral galaxy may feed this way \citep{Czerny:2013aj}.
On the other hand, elliptical galaxies typically lack a substantial inflow owing to a small gas content and low cooling efficiency \citep{Mathews:2003sp}.
Their nuclear star clusters may take over the feeding. The stars shed mass in amounts often large enough to sustain the observed level of AGN activity
\citep{Holzer:1970va,Ciotti:2001nt,Quataertwind:2004,Hopkins:2006mi,Ciotti:2007az,Cuadrawinds:2008,Ho:2009se,Volonteri:2011qk,Miller:2012qu}.
Tidal disruptions \citep{Milosavljevic:2006ti,Macleod:2012kp}, consecutive partial disruptions \citep{Macleod:2013sp}, and stellar collisions \citep{Freitag:2002wj}
account for a small fraction of LLAGN activity, so we ignore such mechanisms. Collisions of ejected stellar winds in the feeding region at $r\sim r_B$ produce
hot gas with a temperature up to $10^7$~K \citep{Lamers:1999,Quataertwind:2004,Cuadrawinds:2008}. The tenuous gas does not cool, but maintains the virial temperature $T_v\sim0.3-1$~keV
and radiates mostly in X-rays. Thus, X-ray studies of LLAGN feeding are warranted.

X-ray studies of nearby LLAGNs include several large \textit{Chandra} projects: an X-ray visionary project (XVP) for Sgr A* (PIs: Baganoff, Markoff, and Nowak) \citep{Wang:2013sc},
an XVP for NGC3115 (PI: Irwin) \citep{Wong:2013ap}, and AMUSE surveys (PIs: Gallo and Treu) \citep{Miller:2012qu}. The unparalleled X-ray spatial resolution of the \textit{Chandra} satellite allows for
the study in unprecedented detail of the gas flow within the BH Bondi radius in several nearby galaxies such as M31, M87, the Milky Way, and NGC3115 \citep{Garcia:2010qg}.
Here we focus on NGC3115, which has an accumulated exposure of $1$~Ms during the year $2012$ with the ACIS-S instrument onboard \textit{Chandra}.
NGC3115 is an S0 lenticular galaxy at a distance of about $d=9.7$~Mpc \citep{Tonry:2001qk}. It host a supermassive BH with mass $M_{BH}\gtrsim10^9M_\odot$
\citep{Kormendy:1996jk,Emsellem:1999qi}. Despite the galaxy being viewed edge-on, the hydrogen column density $N_H$ towards its center is consistent with the local Milky Way $N_H$
\citep{Wong:2011de,Wong:2013ap}. Cold gas is practically absent near the center of NGC3115.  The nucleus has a Bondi radius of $r_B=2-4$~arcsec, which is readily resolved with \textit{Chandra}.
The AGN was only recently found in NGC3115 owing to radio observations \citep{Wrobel:2012zn}. Source radio luminosity is $\nu L_\nu(8.5{\rm GHz})=3.1\times10^{35}{\rm erg~s}^{-1}$.
The nuclear star cluster was extensively observed in the optical band in search
of a supermassive BH with both ground-based instruments \citep{Kormendy:1992ad} and the \textit{Hubble} Space Telescope \citep{Kormendy:1996jk,Emsellem:1999qi}.

The models to study LLAGN feeding have various degrees of complexity. Basic one-zone estimates are typically performed in conjunction with observational studies
to relate the properties of the nuclear star clusters and the observed X-ray emission \citep{Soria:2006ab,Soria:2006bc,Hopkins:2006mi,Ho:2009se,Miller:2012qu,Volonteri:2011qk}.
A more self-consistent approach is to perform radius-dependent modeling. The required radial structure of both the nuclear star clusters and the X-ray emission are available, e.g., for NGC3115.
The radial modeling can quantitatively include a variety of physical effects such as the mass and the energy injection, conduction, and the galactic gravitational potential.
The system of equations can be defined and solved in search for physical solutions \citep{Quataertwind:2004,Shcherbakov:2010cond} encompassing a huge dynamic range
of $\sim10^6$ from the event horizon to far beyond the Bondi radius.  The disadvantages of radial modeling include approximations for vertical flow structure
and the inability to properly deal with the turbulent inhomogeneous medium.  The numerical simulations of the LLAGN feeding allow for the proper treatment of cooling \citep{Gaspari:2013qa},
feedback \citep{Guo:2013ak}, and outflows \citep{Yuan:2012lp,Yuan:2012zz}.  However, full numerical simulations are computationally expensive, which limits the dynamic range and the number
of runs to explore the range of inputs \citep{Yuan:2012lp,Sadowski:2013kc}. In the present paper we adopt radial modeling,
which allows us to compute many solutions and fit the X-ray data in a more consistent way compared to the one-zone estimates. Our results help to illuminate the relative importance
of various physical effects and define the relevant ranges of model parameters such as the BH mass. Our computations provide the starting point
for future numerical simulations of NGC3115 and other LLAGNs.

\begin{figure*}[!hpt]
\plotone{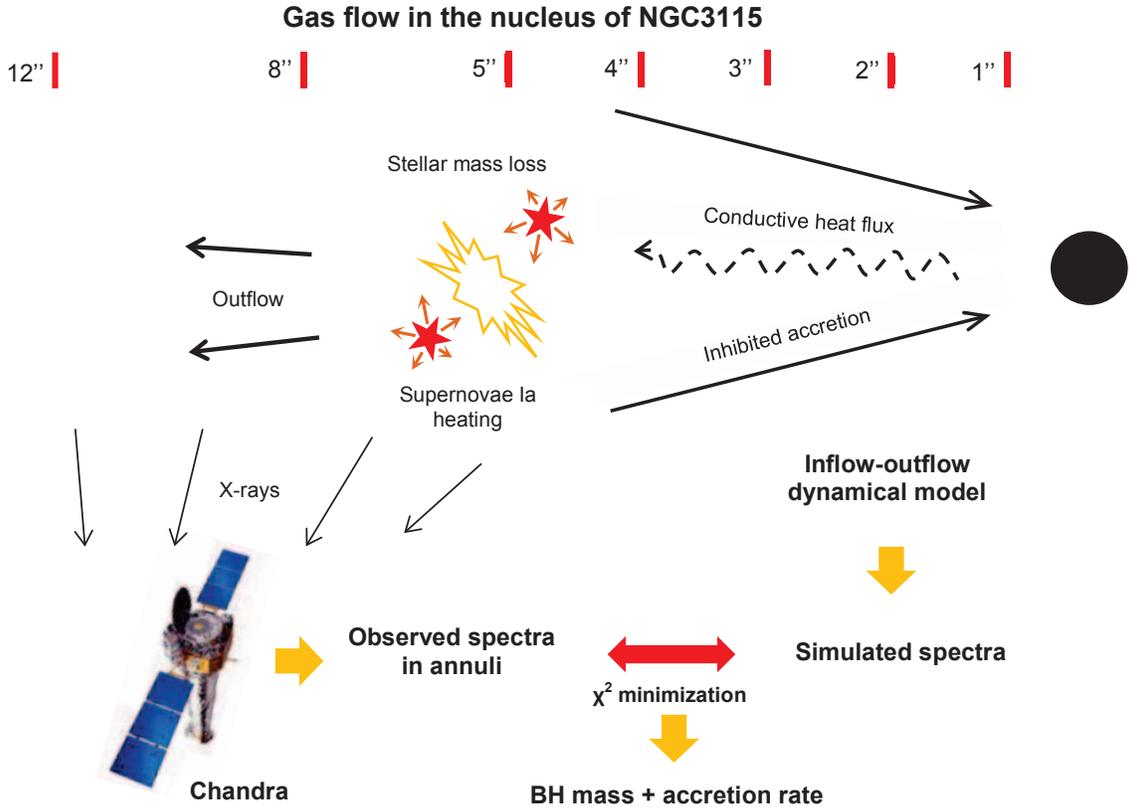}
 \caption{Modeling the hot gas flow onto the supermassive BH in the NGC3115 nucleus. The gas is injected in the feeding region at $r\sim r_B$ by the stars.
Supernovae dominate the energy input in the outer flow $r\gtrsim1$~arcsec, while the collisions of stellar winds dominate the energy input in the inner flow $r\lesssim1$~arcsec.
Most of the gas outflows from the feeding region, while a fraction of the gas accretes.  The extended gas emission is observed in the NGC3115 nucleus
with \textit{Chandra} with the XVP observation (PI: Irwin) and is described in a companion paper \citep{Wong:2013ap}.
We construct radial inflow-outflow solutions for the gas and fit the radius-resolved spectra with simulated spectra.
We find an indication of a low BH mass $<1.3\times10^{9}M_\odot$ and
estimate the mass injection rate normalization, the energy injection rate, and the stagnation point location.
The tick marks, which denote the distance from the BH, are drawn out of proportion. The image of the \textit{Chandra} satellite: copyright NGST.}\label{fig:scheme}
\end{figure*}

The development of such radial gas flow models for NGC3115 and fitting the X-ray XVP data are the topics of this manuscript.
In Section~\ref{sec:stellar_winds} we present the properties of the nuclear star cluster in NGC3115.
We quantify the mass loss by stars, the energy injection by the stellar winds and supernovae, and the angular momentum injection.
In Section~\ref{sec:gas_dynamics} we explore the various effects and features of gas dynamics and devise a radial system of dynamical equations.
We include conduction, the gravitational pull by the enclosed stellar mass, cooling, and the collisional coupling of the ions and the electrons.
In Section~\ref{sec:fitting_data} we outline the procedure of computing radiation from the dynamical gas model and fitting the X-ray data.
We perform optically thin radiative transfer with up-to-date collisional ionization equilibrium (CIE) plasma emissivity and do the radius-resolved spectroscopy.
In Section~\ref{sec:radial_solutions} we present the best-fitting conductive and advective solutions.
We achieve acceptable fits with $\chi^2/\dof=1.00$, which indicates the sufficiency of the radial models. However, we identify room for improvement as prompted by the fit residuals.
In Section~\ref{sec:discussion} we discuss the results and provide conclusions.  The density is found to behave approximately as $n\propto r^{-1}$
over a large dynamic range and across the Bondi radius. We discuss multiple reasons for this density slope.
We identify the limitations of the presented models and discuss directions of future research. The methods of the paper are visualized in Figure~\ref{fig:scheme}.

\section{PROPERTIES OF NUCLEAR STAR CLUSTER}\label{sec:stellar_winds}
The first step to understand LLAGN feeding is to quantify the properties of nuclear star clusters, which provide matter, energy, angular momentum,
and enclosed mass. Nuclear star clusters are ubiquitously present near supermassive BHs \citep{Milosavljevic:2004as,Soria:2006bc,Seth:2008xz,Graham:2009kn,Genzel:2010re}.
The matter, which the cluster stars shed, often constitutes most of the AGN fuel \citep{Ho:2009se}. Luckily, large amounts of data are available on nuclear star clusters
as the by-products of weighing their supermassive BHs, e.g., by the Nuker group (PI: Richstone) \citep{Kormendy:1995,Kormendy:2001al}.
The nucleus of NGC3115 is one of the most studied.  We use the data from earlier ground-based observations by \citet{Kormendy:1992ad},
who modeled the deprojected luminosity and the enclosed mass profiles. Later \textit{Hubble} data showed general agreement with those earlier observations \citep{Emsellem:1999qi}.

\subsection{Enclosed Mass}
A direct by-product of measuring the velocity dispersion is the radial profile of the enclosed mass \citep{Kormendy:1992ad}.
As the surface brightness profile is determined and deprojected, the mass-to-light $M/L$ ratio for the $V$-band is computed at each radius.
A constant value $M/L_V=4.0$ is reached far from the BH, despite the fact that the ratio $M/L_V$ varies close to the BH between different models of the velocity and the mass distributions.
Neglecting any gradient in the stellar mass function, we assume a constant $M/L_V=4.0$ ratio for the enclosed stellar mass at any radius.
Then we multiply the deprojected luminosity profile for the best-fitting D3 stellar dynamic model in \citet{Kormendy:1992ad} by the $M/L_V$ ratio and find the enclosed stellar mass.
Direct inference of the enclosed stellar mass from the velocity dispersion is unreliable near the BH, when the BH mass is not precisely known.
Setting a constant stellar $M/L_V$ allows to disentangle the BH mass from the stellar mass. In the bottom panel of Figure~\ref{fig:feedplot}
we show the computed stellar enclosed mass $M_{\rm enc}$ (solid line) and the total enclosed mass calculated in \citet{Kormendy:1992ad} (dashed line).

\subsection{Mass Injection}\label{subsec:mass_injection}
The most important feature of nuclear star clusters is the ability to inject matter, often in the form of stellar winds, to fuel the supermassive BHs.
Many theoretical and observational studies of the matter injection rates and their relation to the observed quantities were conducted over the years
(see \citealt{Ho:2009se} for a review). The mass loss rate is found to be proportional to the stellar mass and to the stellar luminosity.
The proportionality coefficients depend on the stellar population age $t$. The correlation with stellar mass based on the stellar evolution models is summarized in \citet{Jungwiert:2001}.
Their proposed formula reads
\begin{equation}\label{eq:MdotM}
\frac{\dot{M}_\star}{M_i}\approx \frac{0.055}{t+t_0}
\end{equation} for solar stellar metallicity. Here $M_i$ is the initial stellar mass and $t_0=5$~Myr.
We estimate the stellar age in the nucleus of NGC3115 with the stellar evolution code EzGal \citep{Mancone:2012gj}.
We run the simple stellar population models of \citet{Bruzual:2003op} and \citet{Conroy:2009el,Conroy:2010pk} for solar metallicity
and reach the observed ratio $M/L_V=4.0$ for the age $t\sim5$~Gyr indicative of an old stellar population.
At that age the stellar mass is $M_\star\approx0.62M_i$ \citep{Jungwiert:2001} and the stellar mass loss rate is $\dot{M}_\star=1.8\times10^{-11}M_\star {\rm yr}^{-1}$.
We also discuss the gas and the stellar metallicities in Section~\ref{subsec:gas_metal} below.

Another method to determine the stellar mass loss rate is from the correlation with the source luminosity.
The mass loss rate $\dot{M}$ is proportional to the $V$-band luminosity as
\begin{equation}\label{eq:MdotV}
\dot{M}_\star\approx 3\times10^{-11}\left(\frac{L_V}{L_{\odot,V}}\right){M_{\odot}{\rm yr}^{-1}}
\end{equation} for an old stellar population \citet{Faber:1976jk,Padovani:1993de}. The normalization coefficient is known to within a factor of $2$ \citep{Ho:2009se}.
A similar relation exists between the $\dot{M}$ and the $B$-band luminosity \citep{Ciottiwinds:1991,Athey:2002de}.
The normalizations determined by the formulas~(\ref{eq:MdotM}) and (\ref{eq:MdotV}) agree for NGC3115 to within a factor of $2$.
These formulas are equivalent for a constant adopted $M/L_V$ ratio, and we use the latter one for convenience.
We compute the profile of $\dot{M}_\star$ from the deprojected $V$-band luminosity given by the best-fitting model D3 in \citet{Kormendy:1992ad}.

We present the resultant mass loss rate in the top panel of Figure~\ref{fig:feedplot}.
While the mass loss rate per unit volume sharply rises towards the center, the depicted contribution per unit radius drops inwards.
The area under the curve is the total mass injection rate. For the modeling of gas dynamics we normalize the mass loss rate by
a radius-independent free parameter on the order unity $f_q\sim1$.

\subsection{Energy Injection}\label{subsec:energy_injection}
Several heating mechanisms with comparable power inputs operate in nuclear star clusters.
First, when the mass loss is accomplished via stellar winds, those winds deposit their kinetic energy into the medium.
The collisions of winds turn that kinetic energy into heat. Nuclear star clusters with young stellar populations, such as the one in our Galactic Center,
produce winds with high velocities up to $2,000{\rm km~s}^{-1}$ from Wolf-Rayet and other young stars \citep{Cuadrawinds:2008}.
However, old stellar populations mostly shed matter and produce winds from asymptotic giant branch (AGB) stars \citep{Vassiliadis:1993pa,Hurley:2000fd,Groenewegen:2007xs}
with a correspondent wind velocity under $50{\rm km~s}^{-1}$ \citep{Knapp:1982qw,Marengo:2009ad,Libert:2010wl,Leitner:2011ad}.
We use the terms "stellar winds" and "mass lost by stars" interchangeably, while having in mind that AGB stars shed mass also via planetary nebulae.

The mass-shedding stars move in a combined gravitational field of the BH and the enclosed stellar mass. The velocity of the relative stellar motions
is on the order of the stellar velocity dispersion $\sigma$, which is $\sigma\sim300{\rm km~s}^{-1}$ in the NGC3115 nucleus \citep{Kormendy:1992ad}.
Then the relative stellar motions introduce much more energy than the motions of matter with respect to the injecting stars \citep{Hillel:2013kf}, and the latter is ignored.
For the purpose of the gas dynamical modeling we use the effective stellar wind velocity
\begin{equation}\label{eq:vw_st}
v_{w,\rm st}=c\left(\frac{r_{\rm g}}r\frac{M_{\rm enc}+M_{BH}}{M_{BH}}\right)^{1/2}
\end{equation} given by the Keplerian velocity in the combined gravitational potential. Here $r_{\rm g}=G M_{BH}/c^2$ is the BH gravitational radius.

Another important energy source are supernovae explosions.
According to \citet{Mannucci:2005ja} the supernova rate in S0 galaxies (like NGC3115) is $R_{SN}\sim 4\times10^{-14}{M/M_\odot}{\rm yr}^{-1}$.
Supernovae Type Ia occur in such galaxies more frequently than other kinds due to the large stellar population age.
Each supernova is typically assumed to inject $E_K=10^{51}$~erg of useful energy \citep{Benson:2010jg} into the gas, while the typical ejecta mass is
$M_{\rm ej}\approx1.5M_\odot$ for the Type Ia \citep{Naunberg:1972dw}. Then the specific mass injection rate is $\sim10^{-13}{\rm yr}^{-1}$.
The specific mass injection rate in stellar winds is $\sim10^{-11}{\rm yr}^{-1}$, so that supernovae inject a negligible amount of mass.
The specific energy injection rate in stellar winds is $\sim10^{37}{\rm erg~yr}^{-1}M_\odot^{-1}$ for a typical velocity $v_{w,\rm st}\sim\sigma=300{\rm km~s}^{-1}$.
The specific energy injection rate in supernovae is $\sim4\times10^{37}{\rm erg~yr}^{-1}M_\odot^{-1}$. The supernovae inject more energy than provided by stellar winds.

An important question is whether the energy injection by supernovae may be averaged over the characteristic gas flow timescale.
Since a mass of about $10^{10}M_\odot$ resides at the Bondi radius, one supernova should happen there every $t_{SN}\sim2.5\times10^3$~yrs.
However, the sound crossing time is $t_s\sim 10^6$~yrs. Then about $400$ supernovae happen before the system reacts, so we treat the energy injection from the supernovae on average.
A more detailed a posteriori justification is given in Section~\ref{sec:discussion}.

Some energy is contributed into the feeding region by accreting objects such as low-mass X-ray binaries (LMXBs).
We estimate the mechanical energy output $L_{\rm mech}$ of the accreting objects by equating it to their X-ray luminosity $L_{\rm mech}=L_X$,
which is a natural assumption for the most powerful high efficiency systems. Knowing that $\sim10^4$ photons came from the LMXBs over $1$~Ms \textit{Chandra} observation,
we estimate their X-ray luminosity to be $L_X\sim4\times10^{38}{\rm erg~s}^{-1}$. Then the mechanical luminosity per unit mass is $L_{\rm mech}=10^{36}{\rm erg~yr}^{-1}M_\odot^{-1}$
for a stellar mass of $10^{10}M_\odot$, which is an order of magnitude lower than the energy injection rate in supernovae or stellar winds. We neglect the energy contribution of the LMXBs.

\begin{figure}[h]
\plotone{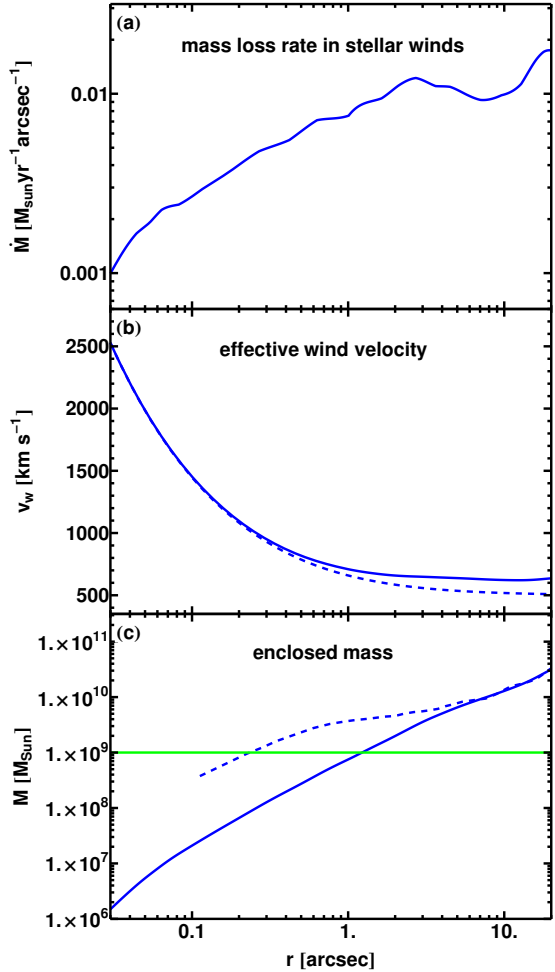}
 \caption{Radial profiles of the quantities in the NGC3115 nuclear star cluster: the mass loss rate $d\dot{M}_\star/dr$ in the upper panel, the effective wind velocity $v_w$ in the middle panel,
 and the enclosed stellar mass $M_{\rm enc}$ in the bottom panel. The area under the curve in the upper panel gives the total mass injection rate $\dot{M}_\star$.
The energy imparted into the stellar winds by the relative stellar motions and the energy injection by the supernovae contribute to the effective wind velocity.
The supernova energy injection rate is equivalent to the energy injection rate of the colliding stellar winds with the velocity $v_{w,SN}$. We set $v_{w,SN}=500{\rm km~s}^{-1}$ in the fiducial model.
The dashed line in the bottom panel corresponds to the total enclosed mass including the BH mass from the D3 model in \citet{Kormendy:1992ad}.
The solid line corresponds to the enclosed stellar mass computed from the surface brightness profile with a constant $M/L_V=4.0$ ratio for the stars.
The horizontal line in the bottom panel represents the BH with a mass $10^9M_\odot$.}\label{fig:feedplot}
\end{figure}

In sum, the two dominant energy contributors are supernovae and colliding stellar winds.
The specific energy injection rate is constant for the supernovae and is a strong function of radius for the colliding stellar winds.
The supernova heating power is equivalent to the power of the colliding stellar winds with a velocity $v_{w,SN}\sim500{\rm km~s}^{-1}$, which we call an effective supernova wind velocity.
We combine the energy inputs into a total effective wind velocity as
\begin{equation}\label{eq:vw_tot}
v_w=\sqrt{v_{w,\rm st}^2+v_{w,SN}^2}=\left(c^2\frac{r_{\rm g}}r \frac{M_{\rm enc}+M_{BH}}{M_{BH}}+v_{w,SN}^2\right)^{1/2}.
\end{equation} In the middle panel of Figure~\ref{fig:feedplot} we plot $v_w$ for the fiducial $v_{w,SN}=500{\rm km~s}^{-1}$ (solid line) and $v_w$
for the same effective supernova contribution, but for the zero enclosed stellar mass $M_{\rm enc}=0$ (dashed line).
In the modeling we leave the effective supernova wind velocity to be a free parameter, but check the best-fitting value of $v_{w,SN}$ for consistency.
\subsection{Angular Momentum Injection}\label{subsec:ang_mom_injection}
\begin{figure}[h]
\plotone{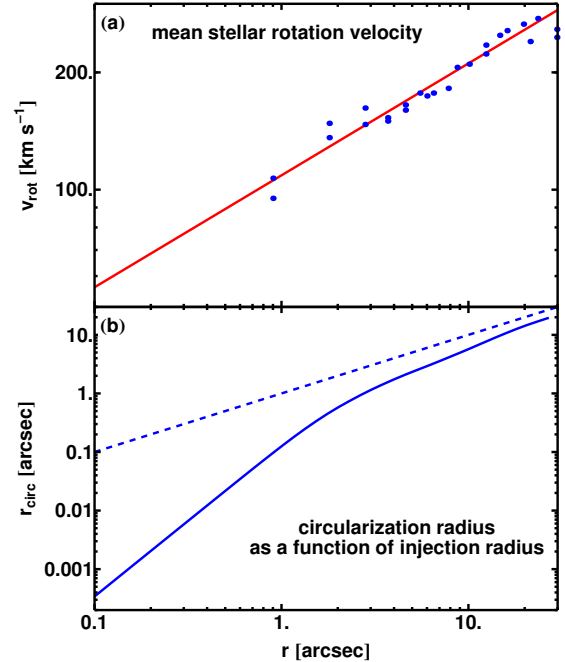}
\caption{Mean angular velocity of the injected stellar winds (top panel) and circularization radius $r_{\rm circ}$ of the matter injected at radius $r$ (solid line in bottom panel).
The fit in the top panel corresponds to $v_\phi=257(r/20)^{0.287}{\rm km~s}^{-1}$ dependence, where $r$ is measured in arcseconds.
The dashed line in the bottom panel corresponds to the equation $r_{\rm circ}=r$, which represents the injection with the Keplerian angular velocity.}
\label{fig:angular}
\end{figure}

The NGC3115 nuclear star cluster possesses a non-zero mean rotation. In their paper \citet{Kormendy:1992ad} report the velocity profiles along the semi-major
and semi-minor axes of the galaxy, which is viewed almost edge-on. The mean rotation is absent along the semi-minor axis, while the mean angular velocity measured along the semi-major axes
reaches more than $200{\rm km~s}^{-1}$. In the top panel of Figure~\ref{fig:angular} we show the radial profile of the mean rotational velocity $v_\phi$ from \citet{Kormendy:1992ad}.
We fit the data points with the power-law
\begin{equation}\label{eq:vphi}
v_\phi=257\left(\frac{r}{20 {\rm arcsec}}\right)^{0.287}{\rm km~s}^{-1},
\end{equation} which quickly approaches zero at a small radius.
In the bottom panel of Figure~\ref{fig:angular} we show the circularization radius $r_{\rm circ}$ as a function of the injection radius $r$ (solid line) given as a solution of the equation
\begin{equation}
r_{\rm circ}~v_K(r_{\rm circ})=r~v_\phi(r),
\end{equation} where the Keplerian velocity is computed in the joint gravitational potential. The dashed line given by the equation $r_{\rm circ}=r$ represents
the injection with the Keplerian angular velocity $v_\phi=v_K$. The presented radial dependence of the angular velocity qualitatively agrees with the transition to a rotationally
supported galactic disk at a large radius and with the transition to purely random stellar motions at a small radius.

\subsection{Gas Metallicity}\label{subsec:gas_metal}
It is difficult to determine the gas metallicity.
The stars in the NGC3115 off-nuclear star clusters have sub-solar metallicity $[Fe/H]\approx-0.5$ close to the center \citep{Arnold:2011oj}.
However, the stellar metallicity might not be a good proxy for the gas metallicity \citep{Su:2013qd}.
The gas metallicity is higher, since only the evolved stars with a large fraction of the heavy elements eject the substantial amounts of mass.
Still, the stellar mass loss rate does not strongly depend on the stellar metallicity of AGB stars \citep{Marigo:2007ju,Weiss:2009ha}.
Since the sound crossing time of the feeding region is only about $t_s\sim10^6$~yrs, the gas metallicity reflects the metallicity of the recently ejected mass.

While the metallicity of the hot gas in NGC3115 cannot be easily measured \citep{Wong:2013ap}, the metallicity of the cooler gas is measured in many other galactic nuclei
to be solar or super-solar \citep{StorchiBergmann:1989pa,Hamann:2002lo}. Relatively few examples exist with sub-solar gas metallicity \citep{Groves:2006we}.
Super-solar metallicity is also favored for the cool absorbing gas near Sgr A* in our Galactic Center \citep{Wang:2013sc}.
With the absence of a better estimate we fix the gas metallicity in NGC3115 at the solar value $Z/Z_\odot=1$.
This approximation is not restrictive. As we discuss in \citet{Wong:2013ap}, the gas metallicity is strongly degenerate with the density normalization.
Since most of the X-rays are emitted in the metal lines, the gas density is inversely proportional to the assumed gas metallicity to preserve the constant density of metals.

\section{GAS DYNAMICS}\label{sec:gas_dynamics}
In Section~\ref{sec:stellar_winds} we characterized the gas injected into the BH feeding region.
In this section we elaborate on the physical laws, which govern the gas dynamics. We first describe the distinct effects and then present a full set of radial equations.
\subsection{Physical Effects}\label{subsec:phys_effects}
\subsubsection{Conduction and Small-scale Feedback}\label{subsubsec:conduction}
Since the early introduction of ADAFs \citep{Bondi1952,Narayan:1995kj}, several effects were shown to break the advective nature
of the hot radiatively inefficient accretion flows and result in a shallow density profile $n\propto r^{-\beta}$ with $\beta=0.5-1.0$, while $\beta=1.5$ for ADAFs.
The flow is not advective, when the energy from the hotter inner flow is deposited into the cooler outer flow, which leads to a super-virial gas temperature.
It immediately follows from the pressure balance equation
\begin{equation}\label{eq:pres_balance}
\frac{1}n\frac{\partial p}{\partial r}=\frac{\partial (n k_B T)}{n \partial r}=-\frac{G M m_p}{r^2}
\end{equation} that in the absence of the source terms that a higher temperature $T$ leads to a shallower density slope $\beta$.

We employ the term ``small-scale feedback" for such energy transfer from the inner flow to the outer flow in an analogy with large-scale feedback, when the central AGN
influences the entire galaxy \citep{Begelman:2005jp,DiMatteo:2005aq,Silk:2010qw}. The two main small-scale feedback processes are convection and conduction.
Convection was shown to be important in collisional flows \citep{Narayan:2000tr,Quataert:2000er}.
Electron heat conduction appears to dominate convection in collisionless flows \citep{Shcherbakov:2010cond}.
No heat is transferred via conduction across magnetic field lines, but the effective conductivity is still high in the turbulent flows as proposed theoretically \citep{Medvedev:2001}
and confirmed with numerical simulations \citep{Parrish:2010lj}. The action of conduction helps to explain the shallow density slope of the Sgr A* accretion flow
\citep{Johnson:2007qw,Shcherbakov:2010cond}.

Outflows may lead to the shallow density profile as well \citep{Yuan:2003sg}. However, it might be non-trivial to disentangle outflows from small-scale feedback.
The simulations by \citet{Yuan:2012lp,Yuan:2012zz} showed both outflows above and below the midplane and convection in the equatorial plane.
Small-scale feedback may help to launch the outflows. When convection or conduction transports energy outwards near the equatorial plane,
outflows are more easily launched above and below the midplane facilitated by the higher gas temperature.
Such a mechanism is distinct from ADIOS \citep{Blandford:1999}, which is based on outflows in the absence of small-scale feedback.
Modeling the flow in one dimension, we do not distinguish between small-scale feedback and outflows.
For the effective combined action of these effects we choose, following \citet{Shcherbakov:2010cond}, unsaturated conduction with the flux proportional
to the temperature gradient $F_{\rm cond}\propto dT_e/dr$ with collisionless conductivity
\begin{equation}\label{conductivity}
\kappa=0.1 \sqrt{k_B T_e/m_e}r n.
\end{equation} The outer flow in NGC3115 is marginally collisional with a mean free path $r_{mfp}\sim0.01r_B$ at the Bondi radius.
However, as we show below, heat conduction effect is subdominant in the feeding region.
The mean free path becomes equal to the radius at $r=r_{mfp}\approx0.1$~arcsec, where conduction becomes dynamically important.
Thus, the solutions computed with high conductivity are physical, and we employ conductivity given by the formula~(\ref{conductivity}) at any radius.
The prescriptions with a lower conductivity in the outer flow reduce the stability of the numerical algorithm, and are avoided.
To test the importance of small-scale feedback, we compute the flow models with and without conduction.

\subsubsection{Gravitational Pull by the Enclosed Stellar Mass}\label{subsubsec:enclosed_mass}
The accretion flows governed by the BH gravity often smoothly connect to the galactic flows governed by the gravity of the enclosed mass \citep{Quataert:2000aj}.
According to Figure~\ref{fig:feedplot}, the enclosed stellar mass in the NGC3115 nucleus exceeds the BH mass at about $r_x\approx1$~arcsec,
which is less than the Bondi radius $r_x\lesssim r_B$. Then, unlike in the Bondi model, constant temperature and constant density are not expected outside of $r_B$.
Outflows need more energy to escape the additional gravitational pull. The gas not bound to the BH may appear bound to the surrounding stellar mass.

\subsubsection{Cooling}\label{subsubsec:cooling}
Cooling is another effect important for AGN feeding. Cool gas readily rushes onto the BH, as it does not have enough pressure to counteract the gravity.
An inverted shape of the cooling curve supports a runaway catastrophe, as the gas loses energy slowly at $T\sim10^6-10^7$~K, but quickly at $T\sim10^5$~K \citep{Sutherland:1993}.
The cooling power is proportional to the density squared $P_{\rm cool}\propto n^2$, and the cooling timescale is inversely proportional to the density $t_{\rm cool}\propto n^{-1}$.
Then this effect is less pronounced in low density systems such as the LLAGNs.

Cooling may influence the accretion in our Galactic Center. The marginal importance of cooling is indicated by \citet{Drappeau:2012dq} close to the plunging region of Sgr A*.
Some models of Sgr A* show runaway cooling in the feeding region \citep{Cuadra:2005ot}, while more realistic models exhibit milder temperature drops \citep{Cuadrawinds:2008}.
A setup very similar to NGC3115 was chosen by \citet{Gaspari:2013qa} for their numerical simulations of accretion flows.
They find only a slight temperature reduction for the low density gas observed in the NGC3115 nucleus. Nevertheless, we include the effect of cooling in the dynamical modeling.
We employ the CIE cooling curve from \citet{Sutherland:1993} and ignore the effects of clumping and spatial inhomogeneity.

\subsubsection{Coupling of Ions and Electrons}\label{subsubsec:coupling}
The thermalization time of a particle distribution in hot tenuous gas is much shorter than the energy exchange time between the electrons and the ions via Coulomb collisions
\citep{Shkarofsky:1966}. We follow the standard practice and consider thermal ions and thermal electrons with temperatures $T_i$ and $T_e$, respectively.
In addition to Coulomb collisions, relatively strong collisionless effects operate at high temperature \citep{Sharma_heating:2007}.
However, we only consider Coulomb collisions in the modeling in the absence of a widely accepted prescription for collisionless coupling.

\subsection{Dynamical equations}\label{subsec:dynamical_eqs}
Following \citet{Shcherbakov:2010cond}, we solve the system of equations on the electron temperature $T_e$, the ion temperature $T_i$,
the electron number density $n=n_e$, and the gas radial velocity $v_r$. The equations are modified from \citet{Shcherbakov:2010cond},
as we ignore the collisionless coupling of the species and the viscous conversion of the gravitational energy into thermal energy.
The latter is justified, because, as we show in Section~\ref{sec:radial_solutions}, the stagnation point in the best-fitting solutions is at $r_{\rm st}\lesssim1$~arcsec.
Then the flow circularization radius lies within $0.05$~arcsec, and the viscous energy production is absent in the observed outer flow.
The thermal energy production via the dissipation of the magnetic field is similarly unimportant till well within the stagnation point \citep{Shcherbakov:2008io}.
Two more modifications are the inclusion of cooling and the inclusion of the galactic gravitational potential. We present the full system of the dynamical equations here.

The mass balance equation is
\begin{equation}\label{eq:mass}
\frac{\partial n}{\partial t}+\frac1{r^2}\frac{\partial(n v_r r^2)}{\partial
r}=\frac{f_q q(r)}{\mu_{av}},
\end{equation} where $\mu_{av}\approx 1.18$ is the average atomic mass per electron for the assumed solar metallicity. The ratio of the number of ions to the number of electrons is
$d=n_{\rm ion}/n=0.91$. We consider the fully ionized species with the relative element abundances given by \textit{wilm} table \citep{Wilms:2000gv}
\footnote{Note, that $\mu_{av}$ and $d$ slightly deviate from those in \citet{Shcherbakov:2010cond} due to their use of a different abundance table.}.
We define the mass source function $q(r)$, such that the mass injection rate plotted in Figure~\ref{fig:feedplot} is
$d\dot{M}_\star/dr=4\pi~r^2~q~m_p$. We normalize $q(r)$ by the dimensionless number $f_q$.
We define the effective isothermal sound speeds
\begin{equation}
c_{se}=\sqrt{\frac{k_B T_e}{m_p}} \quad{\rm and}\quad c_{si}=\sqrt{\frac{k_B T_i}{m_p}}.
\end{equation} Then the Euler equation reads
\begin{equation}\label{eq:Euler}
\frac{D v_r}{D t}+\frac{\partial (n (c_{se}^2+ d~c_{si}^2))}{n \mu_{av}\partial r}+\frac{r_{\rm
g}c^2}{2(r-r_{\rm g})^2}\left(1+\frac{M_{BH}}{M_{\rm enc}}\right)+\frac{f_q q(r)}{n \mu_{av}} v_r=0,
\end{equation} where $D/Dt=\partial/\partial t+v_r \partial/\partial r$ is the Lagrangian derivative.

The relativistic energy exchange rate per unit volume between the electrons and ions via Coulomb collisions is \citep{Stepney:1983dw,Narayan:1995kj}
\begin{eqnarray}\label{coulomb_rel}
F_{ie,{\rm rel}}&=&\frac32\frac{m_e}{m_p}n n_{\rm ion}\sigma_T c \frac{k_B T_i-k_B T_e}{K_2(1/\theta_e)K_2(1/\theta_i)}\ln L \\
&\times&\left[\frac{2(\theta_e+\theta_i)^2+1}{\theta_e+\theta_i}K_1\left(\frac{\theta_e+\theta_i}{\theta_e\theta_i}\right)+
2K_0\left(\frac{\theta_e+\theta_i}{\theta_e\theta_i}\right)\right],
\end{eqnarray} where $K_n$ is the modified Bessel function, the Coulomb logarithm is about $\ln L=20$, and the dimensionless electron and ion temperatures are
\begin{equation}
\theta_e=\frac{k_B T_e}{m_e c^2}, \quad \theta_i=\frac{k_B T_i}{m_p c^2}.
\end{equation} The rate simplifies to
\begin{eqnarray}\label{coulomb_energy}
F_{ie}&=&\frac{3}{2}\sqrt{\frac{2}\pi}\frac{m_e}{m_p}n^2 d \sigma_T c k_B (T_i-T_e)\left(\frac{m_e c^2}{k_B T_e}\right)^{3/2}\ln L \\
&\approx&1.35\times10^{-13} \frac{n^2}{c_{se}^3}(c_{si}^2-c_{se}^2)[{\rm erg~s}^{-1}{\rm cm}^{-3}]\end{eqnarray} in a non-relativistic case.

The energy equations employ the relativistic electron energy
\begin{equation}
u_e \approx \frac32 \frac{0.7+2c_{se}^2m_p/m_e}{0.7+c_{se}^2m_p/m_e}m_p c_{se}^2,
\end{equation} and the CIE cooling power \citep{Sutherland:1993}
\begin{equation}
P_{\rm cool}=\Lambda(T)n^2
\end{equation} for the cooling rate $\Lambda(T)$.
Then the electron energy balance equation is
\begin{eqnarray}\label{eq:energy_electr}
n\frac{D (u_e/m_p)}{D t}-c_{se}^2 \frac{D n}{D t} - \frac{F_{ie}}{m_p}= -\frac{P_{\rm cool}}{m_p}\nonumber\\
+\frac{f_q q(r) (1+d)}{2\mu_{av}}\left(\frac{v_r^2}{2}+\frac{v_w^2}{2}- \frac{5}{2}c_{se}^2\right)
+\frac1{r^2}\partial_r(r^2\kappa\partial_r c_{se}^2),
\end{eqnarray}  where the effective wind velocity $v_w$ is given by the formula~(\ref{eq:vw_tot}).
The ion energy balance is
\begin{eqnarray}\label{eq:energy_prot}
n \frac{D}{Dt}\left(\frac32 c_{si}^2\right)-c_{si}^2\frac{Dn}{Dt}+ \frac{F_{ie}}{m_p}=\nonumber\\
\frac{f_q q(r)(1+d)}{2\mu_{av}}\left(\frac{v_r^2}{2}+\frac{v_w^2}{2}-\frac{5}{2}c_{si}^2\right).
\end{eqnarray} While \citet{Shcherbakov:2010cond} enhanced the rate of Coulomb collisions to enforce the temperature equality $T_e=T_i$ in the feeding region,
we employ the normal rate of Coulomb collisions.

\subsection{Free Parameters and Boundary Conditions}
We search for the stationary solutions of the system of equations~(\ref{eq:mass},\ref{eq:Euler},\ref{eq:energy_electr}, and \ref{eq:energy_prot}) with the shooting method.
The system has four free parameters: the BH mass $M_{BH}$, the normalization of the mass source function $f_q$, the effective supernova wind velocity $v_{w,SN}$,
and the stagnation point radius $r_{\rm st}$. Multiple solutions exist, however, for each set of these parameters.
We identify a set of the natural boundary conditions and the constraints, which leads to a unique solution. These are
\begin{enumerate}
\item{equal electron and ion temperatures $T_i=T_e$ at a large radius,}
\item{the presence of a sonic point in the accretion flow and the absence of shocks,}
\item{and the zero gradient of the electron temperature close to the BH $dT_e/dr=0$.}
\end{enumerate} The third condition is practically equivalent to the requirement of the mere existence of the solution down to the BH horizon.
We also search for advective solutions of the same system of equations by setting the conductivity to zero.
Only the first two conditions are imposed to find a unique advective solution.

\section{FITTING XVP DATA}\label{sec:fitting_data}
Having presented the dynamical model, in this section we discuss the X-ray data and outline the computations of the simulated spectra and the fitting technique.
Previous source modeling relied on earlier \textit{Chandra} observations with $150$~ks total exposure \citep{Wong:2011de}.
New observations with a combined exposure $1$~Ms were performed in 2012 within \textit{Chandra} XVP.
The new data and their model-independent analysis are presented in a companion paper \citet{Wong:2013ap}.
The deep X-ray observations of the NGC3115 nucleus reveal the extended source centered on the BH, which consists of the gas and unresolved point sources.

\subsection{CIE or Non-equilibrium Ionization?}\label{subsec:CIE_NEI}
The gas temperature is about $T=0.3-1$~keV, so that the emission is dominated by metal lines at $E\approx0.8$~keV \citep{Wong:2011de}.
The line emission power is influenced by the gas ionization state.
The collisions of the stellar winds and the shock waves from supernovae lead to the episodes of instantaneous heating.
The heating episodes throw the gas into a non-equilibrium ionization (NEI) state. The CIE is restored after a large number of particle collisions.
The number of collisions is quantified by the ionization timescale, a product of the density by the time $\xi=n t$.
We estimate $\xi$ at a $5$~arcsec radius, where the density is $n=3\times10^{-2}{\rm cm}^{-3}$ \citep{Wong:2011de}.
The region has a sound crossing time $t_s\sim10^6$~yrs, during which about $N\sim400$ supernovae explode. Then the same portion of gas is shocked every $\Delta t=t_s N^{-1/3}\sim10^5$~yrs.
The ionization timescale between shocks is $\xi\sim 10^{11}{\rm s~cm}^{-3}$, for which the flow might not attain full ionization equilibrium \citep{Smith:2010la}.
However, as we show below, cooling is relatively weak in the best-fitting flow solutions. A passage of a single shock might not substantially change the gas temperature,
so that the effective ionization timescale is much larger, and the CIE assumption is justified.
For the gas radiation we use the CIE model \textit{apec} based on ATOMDB 2.0.1 \citep{Foster:2012we} as implemented in XSPEC 12.8 \citep{Arnaud:1996oa}.
The NEI effects are to be explored in future work.

\subsection{Optical Depth Effects}\label{subsec:optical_depth}
The optical depth also influences the line emission power.
Here we show that the X-ray radiation in the NGC3115 nucleus is optically thin to both absorption and resonant scattering.
Let us make a strong assumption that all the X-ray luminosity is concentrated in a single line.
We set the line energy at the peak of the observed gas spectrum $E_{\rm line}=0.8$~keV.
The temperature in the region is $T\sim0.4$~keV \citep{Wong:2011de}, so that the line is subject to thermal broadening by $dv/c\sim6\times10^{-4}$.
Let us now compare the blackbody luminosity in this line with the total observed luminosity to estimate the efficiency of absorption.
The blackbody source function is
\begin{equation}
B_\nu=\frac{2h\nu^3}{c^2}\left(\exp\left(\frac{h \nu}{k_B T}\right)-1\right)^{-1}.
\end{equation} Then the blackbody line luminosity is
\begin{equation}
P_{\rm line}=4 \pi r^2 B_\nu(E_{\rm line})\frac{dv}c \nu_{\rm line} \sim10^{59}{\rm erg~s}^{-1}
\end{equation} emitted by a sphere with a radius $r=1$~arcsec.
This is about $20$ orders of magnitude above the observed luminosity.

Resonant scattering may have a larger effect as it is found to change the surface brightness profiles of elliptical galaxies \citep{Shigeyama:1998aj}.
\citet{Shigeyama:1998aj} estimate the emission averaged optical depth to be around $\tau_{\rm sc}\sim5$ over the scattering column density $N_{\rm sc}\sim3\times10^{20}{\rm cm}^{-2}$ for the relevant
gas temperature. The scattering column density to the center of NGC3115 is about $N_{\rm sc}\sim3\times10^{18}{\rm cm}^{-2}$ and the correspondent optical depth is $\tau_{\rm sc}\sim0.05\ll1$.
Thus, the gas emission is optically thin to both absorption and scattering.

\subsection{Point sources}\label{subsec:point_sources}
The contamination by the point sources complicates the modeling of the gas emission.
We subtract the brightest isolated objects, but source confusion precludes the reliable subtraction at radii $r\lesssim4$~arcsec.
The weaker and confused point sources contribute to the extended emission.
A reliable spectral model is the key to discriminate such emission from the gas emission.

LMXBs comprise most of the resolved and some of the unresolved point source emission.
The combined spectrum of the resolved LMXBs is an absorbed power-law with index $\Gamma_{LMXB}=1.61$ \citep{Wong:2011de}, and we use the same index to model the unresolved LMXBs.
The nuclear LMXB luminosity is proportional to the stellar mass \citep{Gilfanov:2004qe,Miller:2012qu}.
However, as the nuclear luminosity in NGC3115 is dominated by a few bright sources, the Poisson noise in the proportionality coefficient is large.
We leave the normalization of the LMXB luminosity to be a free parameter.

Cataclysmic variables and coronally active stars contribute to the diffuse X-ray emission as the so-called CV/AB component \citep{Revnivtsev:2006da,Revnivtsev:2008kd}.
Following \citet{Wong:2013ap} we model the CV/AB contribution as an absorbed sum of a power-law with an index $\Gamma_{CV}=1.915$ and a thermal component with $T_{CV}=0.763$~keV.
The total CV/AB luminosity is computed from the $L_K-L_X$ relation and the surface brightness is taken to be proportional to the optical surface brightness.
Each CV/AB source is relatively weak, so that many sources contribute to the emission, and Poisson noise is insignificant.
The CV/ABs is a sub-dominant component in the inner flow \citep{Wong:2013ap}.

\subsection{Point Spread Function}\label{subsec:PSF}
The \textit{Chandra} observations probe the inner several arcseconds around the supermassive BH in NGC3115.
Since spatial variations are expected on the scale of $\lesssim1$~arcsec, the results of such observations are affected by photon redistribution
due to the finite size of the point spread function (PSF).
An implementation of the resultant PSF spreading is generally available in XSPEC as the mixing models, yet no such model exists for the \textit{Chandra} PSF.
We implement the PSF spreading in \textit{Mathematica}~9 and perform consistency checks.
For simplicity, we adopt an energy-independent Gaussian PSF with width $\sigma_{PSF}=0.27''$, which fits the core of the surface brightness profile of a nearby point source.

\subsection{Procedure}\label{subsec:procedure}
There are many ways to compare the simulated emission from the accretion flow model to the observations.
For example, \citet{Shcherbakov:2010cond} compared the energy-integrated profiles of the surface brightness.
This approach introduces a degeneracy between the temperature and the density: the high temperature low density model produces the same surface brightness as the low temperature high density model.
The degeneracy may be broken with the use of the spectrum.
Fitting the radius-integrated spectrum \citep{Wang:2013sc} one obtains the relative contributions of gas at the different temperatures with little information about the spatial distribution.
Thus, we maintain both the spectral and the spatial information, while comparing the simulated emission to the data.

Following \citet{Wong:2013ap} we divide the BH feeding region into circular rings centered on the BH.
In this paper we limit ourselves to an outer radius of $12$~arcsec, which is far outside of $r_B$.
We define $7$ rings with the projected radii in the ranges $0-1$~arcsec, $1-2$~arcsec, $2-3$~arcsec,  $3-4$~arcsec, $4-5$~arcsec, $5-8$~arcsec,  and $8-12$~arcsec.
The spectrum of each ring is extracted and grouped with a minimum of $25$ photons per bin for a total of $247$ bins over the $7$ rings.
Having defined the observed spectra, we calculate the simulated spectra and compute the chi-square statistic.

The simulated spectrum in each ring is the sum of the fixed CV/AB contribution, the fixed background, the power-law LMXB component with a free normalization,
and the gas component. Since the background dominates at high energies, we set the high energy limit at $E_{\max}=6$~keV.
We set the low energy limit at $E_{\min}=0.5$~keV as the lowest bin energy for the grouped observed spectrum.

The gas properties are defined by the computed profiles of the temperatures $T_i(r)$ and $T_e(r)$ and the electron density $n(r)$ for the radius from $r_{\min}=2000r_{\rm g}=4\times10^{-3}$~arcsec
to $r_{\max}=7.8\times10^6r_{\rm g}=16$~arcsec.  We set the density to zero outside of this radial range. We calculate the simulated spectra fully self-consistently.
We divide the flow into many spherical shells and compute the contributions of each shell into the projected rings, while taking into account the PSF spreading.
We find a joint $\chi^2$ as a sum over all $7$ rings. We perform the steepest descent search for a minimum of $\chi^2$ over the set of the model parameters.
We explore the BH masses in the range $M_{BH}=(0.7-2.0)\times10^9M_\odot$ motivated by the dynamical modeling of the stellar motions \citep{Kormendy:1996jk,Emsellem:1999qi}.
We do not restrict the other three free parameters $f_q$, $v_{w,SN}$, and $r_{\rm st}$. We find the best-fitting conductive and advective solutions.

\section{RADIAL INFLOW-OUTFLOW SOLUTIONS}\label{sec:radial_solutions}
\subsection{Solutions with Conduction}\label{subsec:solutions_conduction}
\begin{figure}[h]
\plotone{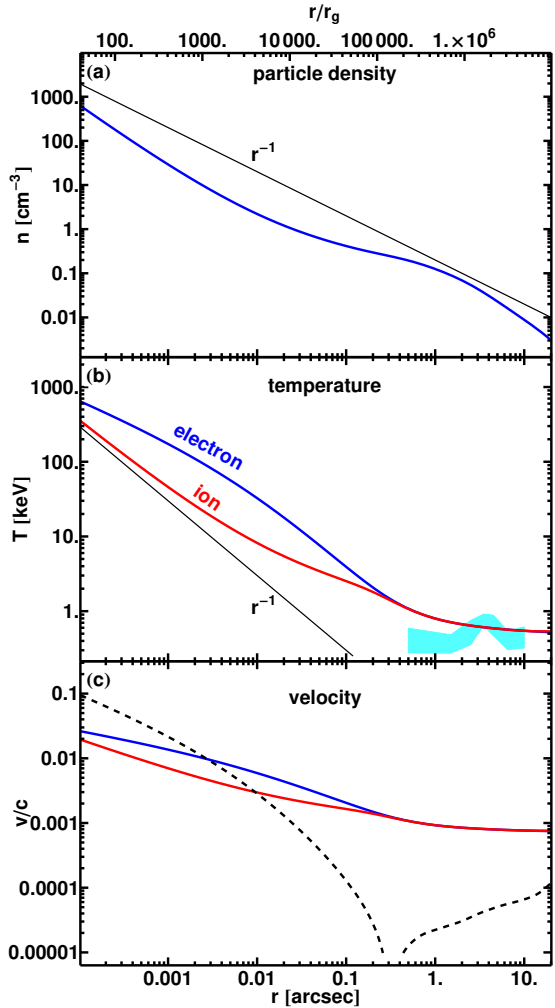}
\caption{Dynamical quantities in the best-fitting solution with conduction: density (top panel), temperature (middle panel), and velocity (bottom panel).
The thin lines in the top and the middle panels show the $r^{-1}$ power-law for comparison. A filled area in the middle panel designates the $90\%$ confidence range of the gas temperature
for the "single-T \textit{apec} per annulus" model. The blue/top solid line in the bottom panel shows the effective isothermal electron sound speed $c_{se}$,
the red/bottom solid line shows the effective isothermal ion sound speed $c_{si}$, and the dashed line shows the absolute value of the gas radial velocity $|v_r|$.}
 \label{fig:solution}
\end{figure}
The dynamical structure of the best-fitting solution with conduction is shown in Figure~\ref{fig:solution}.
This solution is achieved at the lower BH mass boundary $M_{BH}=0.7\times10^9M_\odot$ for the mass loss rate normalization $f_q=0.150$,
the effective supernova wind velocity $v_{w,SN}=521.6{\rm km~s}^{-1}$, and the stagnation point radius $r_{\rm st}=0.33$~arcsec.
It reaches $\chi^2/\dof=1.001$ for $\dof=236$ and has an accretion rate $\dot{M}\approx2\times10^{-4}M_\odot {\rm yr}^{-1}$.
This accretion rate is a factor of $100$ lower than Bondi accretion rate \citep{Wong:2011de}.
However, the correspondent accretion power $\dot{M}c^2\sim10^{43}{\rm erg~s}^{-1}$ is still about $7.5$ orders of magnitude larger than the observed jet radio power.
The density (shown in the top panel) behaves approximately as $n\propto r^{-1}$ over the large range of the radius.
The density does not flatten out outside of the Bondi radius. The electron temperature is higher than the ion temperature due to heat conduction,
which primarily influences the electrons. The ion temperature is approximately virial at all radii, which corresponds to $T_i\propto r^{-1}$ in the inner flow.
The slope of $T_i$ is relatively flat in the outer flow, where the enclosed mass increases with radius as $M_{\rm enc}\propto r$, so that $T_i\propto M_{\rm enc}r^{-1}\propto r^0$.
The gas inflow velocity (shown in the bottom panel) exceeds the sound speed at a relatively large radius $r\sim10^3r_{\rm g}$.
This is a consequence of the rising electron heat capacity \citep{Shcherbakov:2008io} and the absence of super-virial heating.
The outflow velocity is much below the sound speed in the outer flow, so that the outflow is subsonic.
The determined stagnation point radius corresponds to the circularization radius of $r_{\rm circ}=2500r_{\rm g}=5\times10^{-3}{\rm arcsec}$ according to Figure~\ref{fig:angular}.
Relatively little X-ray emission originates within this radius in a non-cooling accretion flow.

\begin{figure}[h]
\plotone{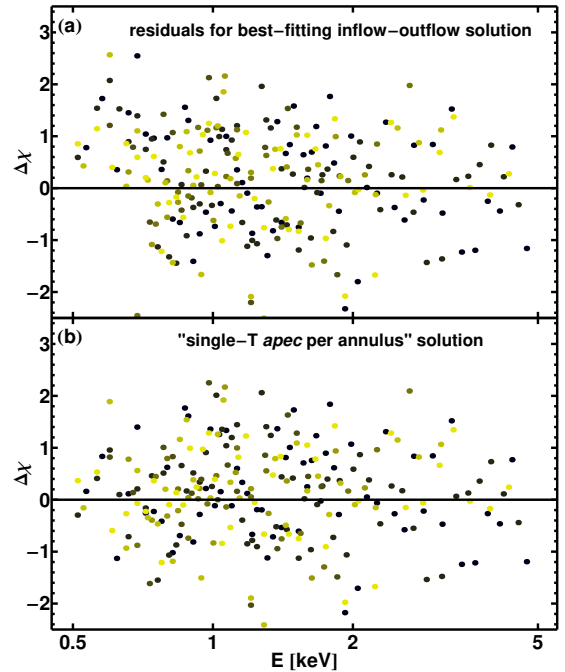}
 \caption{Fit residuals for the best-fitting inflow-outflow solution with conduction (top panel) and for the "single-T \textit{apec} per annulus" model (bottom panel).
 The best-fitting conductive solution reaches $\chi^2/\dof=1.001$ for $4$ free gas parameters, while the "single-T \textit{apec} per annulus" model reaches $\chi^2/\dof=0.895$
for $14$ free gas parameters. The residuals for the inner annulus are shown as the blue/dark dots, for the outer annulus as the green/light dots, and
for the intermediate annuli as the dots of the intermediate colors/shades of grey. The best-fitting conductive solution undepredicts the observed soft flux as evident
from the systematic trend at the lowest energies in the top panel.}
\label{fig:residuals}
\end{figure}

\begin{figure}[h]
\plotone{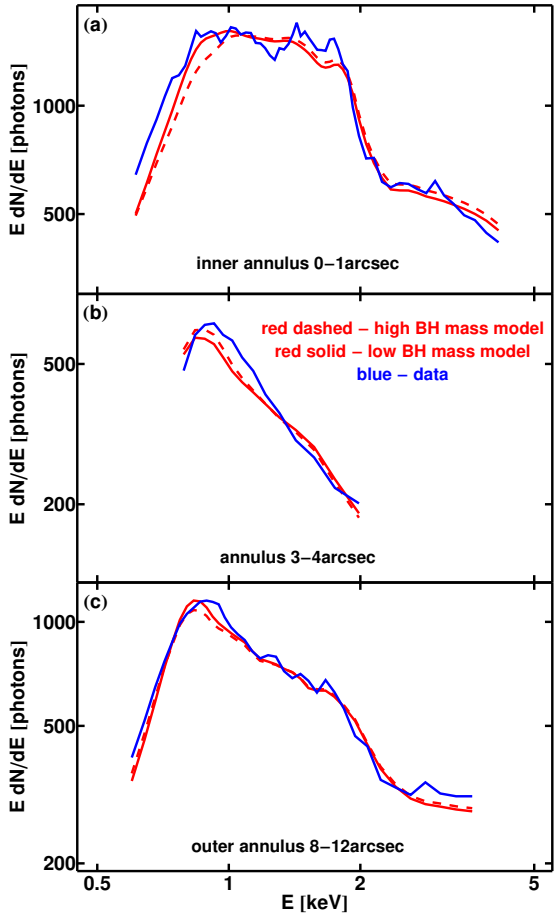}
 \caption{Low BH mass model (in solid red/light), high BH mass model (in dashed red/light), and data (in solid blue/dark) for the best-fitting inflow-outflow models with conduction:
 in the annuli with projected radii within $0-1$~arcsec range (top panel), $3-4$~arcsec range (middle panel), and $8-12$~arcsec range (bottom panel).
 The spectra are smoothed over $5$ adjacent energy bins to lower photon noise. Low BH mass model is computed for $0.7\times10^9M_\odot$ BH, while high BH mass model has a $1.8\times10^9M_\odot$ BH.}
\label{fig:model_data}
\end{figure}

In the middle panel of Figure~\ref{fig:solution} we depict the $90\%$ confidence range of the temperature (green/light area) in the "single-T \textit{apec} per annulus"
best-fitting model presented in \citet{Wong:2013ap}. In this model the observed spectrum in each annulus is fitted independently
with a single-temperature \textit{apec} component instead of drawing the gas temperatures from a smooth radial profile.
The point source and the background contributions are computed the same way in both kinds of models.
The temperature in the "single-T \textit{apec} per annulus" best-fitting model agrees with the temperature in the best-fitting conductive solution
at large radii $r>2$~arcsec,  but deviates down in the inner flow.

As we show in Figure~\ref{fig:residuals} the "single-T \textit{apec} per annulus" model reaches lower $\chi^2/\dof=0.895$ with the more uniform residuals $\Delta(\chi)$.
The best-fitting solution with conduction underpredicts the observed soft flux as evident from the systematic trend at the lowest energies in the top panel.
In sum, while our solution could be improved with the lower temperature in the inner flow, it already provides an acceptable fit to the data with $\chi^2/\dof=1.001$.
The reasons for this underprediction are explored in \citet{Wong:2013ap}, the main hypothesis being the presence of an inner soft component, which appears to be extended.
This soft component could either be a cool diffuse gas or a distinct population of point sources. \citet{Wong:2013ap} develop a two-component gas model and obtain a temperature profile
of the hotter component, which agrees with the best-fitting profile of the electron temperature.
The data and the model are shown for selected annuli in Figure~\ref{fig:model_data} for low BH mass ($0.7\times10^9M_\odot$) and high BH mass ($1.8\times10^9M_\odot$)
best-fitting solutions with conduction. Lack of strong soft emission is evident in both the inner and the outer annuli for both BH masses.
Substantial non-thermal emission convolved with \textit{Chandra} response function is responsible for $1.9$~keV bump in the outer annuli, while the extended bump in $1.3-1.9$~keV energy range
in the inner annuli is mainly emitted by the hot inner accretion flow.
The annuli with projected radii within $3-4$~arcsec exhibit relatively soft spectrum, while the models are substantially harder. As indicated by "single-T \textit{apec} per annulus" model,
a hotter thermal plasma provides a better fit to that spectrum. Since the galactic gravitational potential dominates the BH potential at $\gtrsim1$~arcsec distance, then the differences between
the low BH mass and the high BH mass models are the most evident in the inner annulus. The high BH mass model has a higher virial temperature.
This leads to further underprediction of the soft flux emitted by the inner accretion flow, so that high BH mass model provides a worse fit to the data.

\subsection{Advective Solutions and Comparison}\label{subsec:solutions_advective}
\begin{figure}[h]
\plotone{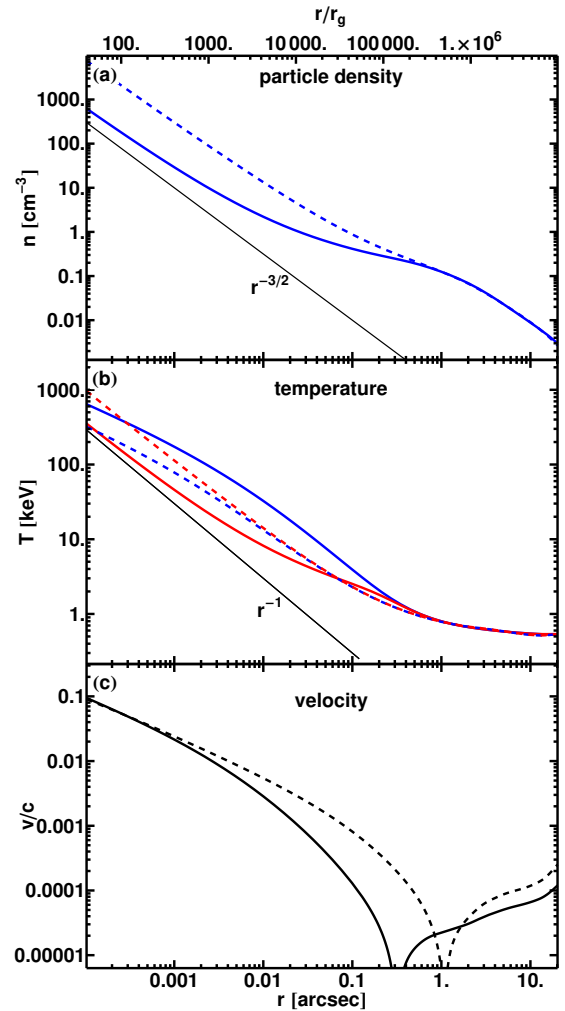}
\caption{Comparison of the dynamical quantities in the best-fitting solutions with and without conduction:
density (top panel), temperature (middle panel), and radial velocity (bottom panel). We show the quantities in the conductive solution (thick solid lines), the quantities
in the advective solution (thick dashed lines), and the simple power-laws (thin lines).
Shown in the middle panel are the electron temperature $T_e$ (blue/upper solid line) and the ion temperature $T_i$ (red/lower solid line) for the solution with conduction,
the electron temperature $T_e$ (blue/lower dashed line) and the ion temperature (red/upper dashed line) $T_i$ for the advective solution.
Note that $T_e>T_i$ in the conductive solution, while $T_e<T_i$ in the advective solution. The density is shallower in the flow with conduction.
Both density profiles asymptote to $n\propto r^{-3/2}$ and both ion temperature profiles asymptote to $T_i\propto r^{-1}$ in the inner flow.
The conductive solution has the sonic point closer in, while the inner asymptotic velocities are comparable in two models.}
 \label{fig:comparison}
\end{figure}
We explore not only the solutions with conduction, but also the advective solutions, where the conductivity is set to zero.
The comparison between these cases helps to explore the role of conduction. In Figure~\ref{fig:comparison} we show the dynamical quantities
for the best-fitting solution without conduction (dashed) and for the best-fitting solution with conduction (solid).
The best fit among the advective solutions is also achieved at the lower BH mass boundary $M_{BH}=0.7\times10^9M_\odot$.
The correspondent values of the free parameters are $f_q=0.289$, $v_{w,SN}=510.0{\rm km~s}^{-1}$, and $r_{\rm st}=1.08$~arcsec.
This solution reaches $\chi^2/\dof=0.998$ and has an accretion rate $\dot{M}\approx2\times10^{-3}M_\odot {\rm yr}^{-1}$.
The values of the free parameters are similar in the best-fitting conductive and advective solutions, except the stagnation point is much further out in the advective solution
and the accretion rate is much higher. This difference is a natural consequence of conduction.
The density profiles in both best-fitting solutions asymptote to the steep Bondi profile $n\propto r^{-3/2}$ in the inner flow.
However, the inner flow density and the accretion rate are a factor of $10$ higher in the advective solution.
This factor may be even larger, when super-virial heating is included \citep{Johnson:2007qw,Shcherbakov:2010cond}.
As super-virial heating is likely important in the inner flow, the computed accretion rate is an upper limit on the rate of mass crossing the event horizon.
The electron temperature in the advective solution is lower than the ion temperature $T_e<T_i$ due to cooling in the outer flow
and the higher electron heat capacity in the inner flow. The solution with conduction has a factor of $3$ lower inner ion temperature.
The energetics of the outer flow are mainly determined by the mass injection and the energy injection, so that the properties of the outer flow are similar between the two best-fitting solutions.

\begin{figure}[h]
\plotone{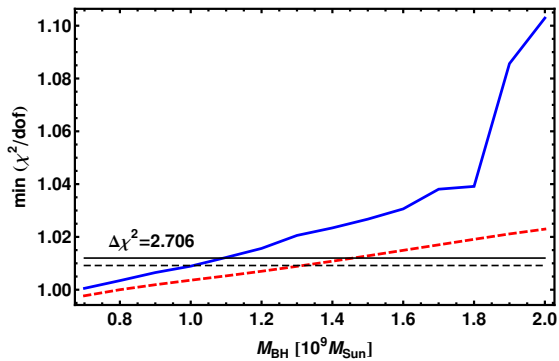}
\caption{Minimum $\chi^2/\dof$ as a function of the BH mass: for the inflow-outflow model with conduction (blue/solid line)
for the advective inflow-outflow model (red/dashed line). A smaller BH mass is preferred in both types of models.
We only explore the BH mass above $0.7\times10^9M_\odot$ as consistent with the dynamical models of stellar motions.
The $90\%$ confidence lines with $\Delta\chi^2=2.706$ are shown for the models with conduction (thin solid) and the advective models (thin dashed).}
 \label{fig:BHmass}
\end{figure}
One of the most important results of the presented model fitting is the BH mass.
In Figure~\ref{fig:BHmass} we show the reduced $\chi^2$ as a function of the BH mass for both the best-fitting advective models (red/dashed line) and
the best-fitting models with conduction (blue/solid line). Both types of models show a clear rising trend with the BH mass
in agreement with Figure~\ref{fig:model_data} and discussion of the spectral features therein.
The $90\%$ confidence range limits the BH mass to below $1.1\times10^9M_\odot$ for the conductive models and to below $1.3\times10^9M_\odot$ for the advective models
in agreement with the latest BH mass estimates \citep{Emsellem:1999qi}. However, the adopted modeling has many caveats,
which should be carefully examined before the firm conclusions are drawn about the BH mass. Here we demonstrate that it is possible to discriminate
between the models with the different BH masses by fitting the X-ray data. We discuss the caveats of the modeling in the next section.

\section{DISCUSSION AND CONCLUSIONS}\label{sec:discussion}
\subsection{Summary of Results}\label{subsec:summary}
In the paper we present the modeling of the X-ray data from $1$~Ms \textit{Chandra} XVP observation of the NGC3115 center.
We connect the properties of the nuclear star cluster known from optical observations to the properties of the X-ray emitting hot gas.
We construct the radial inflow-outflow dynamical models, which include many physical effects: the matter and the energy injection by stellar winds and supernovae,
conduction, the additional gravitational pull by the enclosed mass, cooling, and Coulomb collisions.
We simulate the X-ray emission from the models and fit the set of the X-ray spectra in concentric annuli around the BH.
We find best-fitting models with an acceptable $\chi^2/\dof\approx1.00$. The proposed models are sensitive to the BH mass and favor low values $<1.3\times 10^9M_\odot$.
We estimate the normalization of the mass source function to be $f_q\approx0.15-0.30$, which is somewhat smaller than the expected value $f_q=1$.
We discuss below the reasons for the deviation of $f_q$ from unity distinct from the uncertainties in the mass loss rate.
The best-fitting effective supernova wind velocity is $v_{w,SN}\approx521{\rm km~s}^{-1}$, which corresponds to the rate of the supernova explosions
$R_{SN}\sim 3\times10^{-14}{M/M_\odot}{\rm yr}^{-1}$ for the fiducial energy release $E_K=10^{51}{\rm egs~s}^{-1}$ per event.
This estimated event rate is consistent with the observed event rate in S0 galaxies like NGC3115 \citep{Mannucci:2005ja}.
The stagnation point is at $r_{\rm st}\approx0.33$~arcsec for the best-fitting solution with conduction and at $r_{\rm st}\approx1.08$~arcsec for the best-fitting advective solution.
Therefore most of the "accretion flow" seen by \textit{Chandra} is outflowing from the region, while the stagnation radius scale is barely resolved.
We find that the best-fitting conductive and advective solutions behave similarly in the outer flow, yet the advective solution has
the higher density and the lower electron temperature in the inner flow.

\begin{figure}[h]
\plotone{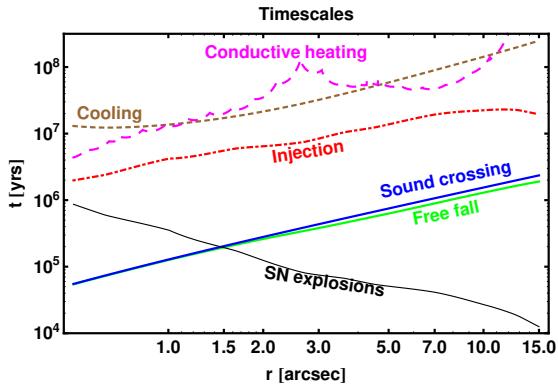}
 \caption{Timescales as a function of radius: the sound crossing time $t_s$ (blue/upper thick solid line),
 the free-fall time $t_{ff}$ in the joint gravitational field (green/lower thick solid line), the cooling time $t_{\rm cool}$ (brown/short-dashed line),
 the conductive heating time $t_{\rm cond}$ (magenta/long-dashed line), the matter injection time $t_{\rm inj}$ (red/dot-dashed line),
 and the time between consecutive supernovae $t_{SN}$ (black/thin solid line). Cooling is unimportant as the cooling time is about $100t_{ff}$.
 Conductive heating is unimportant in the outer flow as well due to large $t_{\rm cond}$. The relatively long mass injection timescale signifies
 the slow gas velocity $v_r$ compared to the sound speed $c_s$.}
\label{fig:timescales}
\end{figure}

It is instructive to compare the relative strengths of the various effects by computing the correspondent timescales.
In Figure~\ref{fig:timescales} we plot the timescales in the feeding region. The sound crossing time (blue/upper thick solid line)
\begin{equation}
t_s=\sqrt{\frac35}\frac{r}{c_{se}}
\end{equation} is about the free fall time (green/lower thick solid line)
\begin{equation}
t_{ff}=\frac{r}c\sqrt{\frac{r}{r_{\rm g}}\left(1+\frac{M_{\rm enc}}{M_{BH}}\right)^{-1}},
\end{equation} so that the gas temperature is close to virial at any radius.
We also plot the cooling time (brown/short-dashed line)
\begin{equation}
t_{\rm cool}=\frac{u_e n}{P_{\rm cool}},
\end{equation}
the conductive heating time (magenta/long-dashed line)
\begin{equation}
t_{\rm cond}=\frac{u_e n}{Q_+}, \quad\text{where}\quad Q_+=-{\rm div}(F_{\rm cond})
\end{equation} is the divergence of the conduction heat flux, and the mass injection time (red/dot-dashed line)
\begin{equation}
t_{\rm inj}=\frac{n}{f_q q}.
\end{equation}
The cooling timescale is about $100$ times the free-fall timescale $t_{\rm cool}/t_{ff}\sim100$, so that cooling is expected to be unimportant
in the modeled hot-phase gas in the non-rotating flow \citep{Gaspari:2013qa}.
The mass injection time is much larger than either $t_s$ or $t_{ff}$, which indicates a relatively slow radial gas velocity $v_r\sim0.1c_s$.
The conductive heating time is very large outside of $r\sim1$~arcsec, so that the solutions with and without conduction behave similarly in the outer flow.
This timescale gets comparable to the injection time inside of $r\sim1$~arcsec, which suggests the importance of conduction at those radial scales.
The comparison of timescales also helps to establish the self-consistency of the model. The time between consecutive supernova explosions (black/thin solid line)
\begin{equation}
t_{SN}=\frac{2E_K}{4\pi r^3f_q q m_p v_{SN}^2}
\end{equation} becomes longer than the sound crossing time at $r\sim1$~arcsec.
Thus, the energy injection from the supernovae cannot be treated on average in the inner flow. However, this mechanism is subdominant at $r\lesssim1$~arcsec
as evident from the middle panel in Figure~\ref{fig:feedplot}: the collisions of stellar winds supply most of the energy in the inner flow.
The inner accretion flow experiences relatively weak disturbances from supernova explosions, which are washed away on the dynamical timescale.
Then the average energy injection rate is well-defined at any radius.

\subsection{What Does the Density Slope Mean?}\label{subsec:density_slope}
The modeling of the resolved X-ray emission gives the gas density slope. We find the shallow density profile $n\propto r^{-\beta}$ with $\beta\approx1$
across a large range of scales in the NGC3115 nucleus. As briefly discussed in Section~\ref{subsec:phys_effects}, the shallow density profile commonly occurs in CDAFs
\citep{Quataert:2000er,Narayan:2000tr}, in the accretion flows with conduction \citep{Johnson:2007qw,Shcherbakov:2010cond}, and in the accretion flows with the outflows above and below the midplane
\citep{Blandford:1999,Yuan:2003sg,Yuan:2012lp}. However, there are other reasons to have $\beta\approx1$ near the Bondi radius in the hot gas flows.
Let us elaborate on the effects either directly responsible for the shallow density profile or calling for the extension of the aforementioned explanations.

First, we examine the original Bondi solution as computed by \citet{Bondi1952}. Their curve II in Figure~5 shows the relevant case of an adiabatic transonic inflow for the
adiabatic index $\Gamma=5/3$.  In this solution the density slope $\beta$ is a function of radius and changes from $\beta=0$ at $r\gg r_B$ to $\beta=1.5$ at $r\ll r_B$.
The steep asymptotic behavior $n\propto r^{-3/2}$ is  achieved only very deep inside the Bondi sphere.
The Bondi flow has $\beta=1.0$ at a tenth of the Bondi radius $r=0.1r_B$, which corresponds to their dimensionless radius $x=0.2$.
The slope at a radius $r=0.5$~arcsec probed by the \textit{Chandra} satellite in NGC3115 is expected to be $\beta<1$ even in a fully advection-dominated Bondi-like flow.

Second, we model the material in NGC3115 to outflow from the stagnation point at $r_{\rm st}\lesssim1$~arcsec.
However, when the mean radial velocity is much smaller than the sound speed $v_r\ll c_s$, then the small-scale feedback and the outflows have the same power for both positive and negative $v_r$.
The pressure balance is practically hydrostatic for small $v_r$ and is given by the equation~(\ref{eq:pres_balance}).
Then small-scale feedback and outflows make the density profile shallow in both the inflow region and the outflow region.
However, the density in the outflow asymptotes to $n\propto r^{-2}$, if the radial velocity is large, while the density is constant $n={\rm const}$ in the Bondi inflow.
Models with large outflow velocity $v_r\sim c_s$ are disfavored for NGC3115, while being viable for Sgr A* \citep{Quataertwind:2004}.

Third, continuous mass injection modifies the mass conservation law, so that $\dot{M}\propto n v_r r^2\ne{\rm const}$.
Mass injection is the dominant term in the density balance at radii $0.5-10$~arcsec in the best-fitting solution with conduction.
Then mass conservation law is inapplicable in the feeding region of NGC3115 near the Bondi radius $r\sim r_B$. The density slope in the outer flow is influenced by the matter source term.

Fourth, the region near and outside of the Bondi radius $r\gtrsim r_B$ is influenced by the gravity of the enclosed mass.
The correspondent gravitational potential does not flatten, but increases with radius. Then the gas outflow velocity stays small and the asymptotic outflow behavior is not reached.
The virial temperature is much higher in the outer gas, when the enclosed mass from the nuclear star cluster is included.
As the gas temperature closely follows the virial temperature, the change in the virial temperature profile influences the density profile.
The density profile $n\propto r^{-1}$ is commonly observed in the hot flows outside of the Bondi radius \citep{Allen:2006fw,Wong:2011de}, where the galactic gravitational potential matters.

The interplay of these four processes and effects determines the gas density profiles in the LLAGNs.
Despite the slope $n\propto r^{-1}$ approximates the density profile in NGC3115 over a large radial range,
the local slope $-d\log n/d\log r$ at a given $r$ often substantially deviates from $\beta=1$. The absence of a single behavior
over a large dynamic range demotivates us from isolating the self-similar solutions.

\subsection{Limitations of the Dynamical Model}\label{subsec:limitations_model}
Despite being able to fit the data, the presented models are not fully self-consistent.
Let us examine the drawbacks and the limitations of the models and outline a more self-consistent treatment.

\subsubsection{Inhomogeneous Medium}\label{subsubsec:inhomogeneous}
The observational studies of Sgr A* suggest inhomogeneous gas near the Bondi radius \citep{Baganoff:2003,Muno_diff:2008,Wang:2013sc}.
Regions with vastly different temperatures readily co-exist, while the \textit{Chandra} satellite only sees the hot dense counterparts with temperature $T>0.3$~keV.
Nevertheless, the observed gas temperature $T=0.3-1$~keV in NGC3115 agrees well with the virial temperature $T\sim T_v$, and the observed density is reproduced with
the normalization of the mass source function $f_q$ on the order unity. Then the hot gas likely constitutes the dominant gas component.

The filling factor of this hot component may still be below unity $f_V<1$. In this case the mean density required to reproduce the observations is lower.
More generally, gas with lower mean density reproduces the observations, when substantial density fluctuations are present.
Since the emissivity is proportional to $n^2$, then thinking of the best-fitting density as the root-mean-squared quantity effectively takes the inhomogeneities into account.

\subsubsection{Non-stationary Solutions}\label{subsubsec:non-stationary}
A wide range of non-stationary behaviors, such as oscillation cycles, may occur in accretion flows.
The best-fitting energy of the outflowing gas is barely enough to escape the gravitational potential of the enclosed mass.
The temperatures in the best-fitting "single-T \textit{apec} per annulus" model are even lower \citep{Wong:2013ap}, so that the gas may be unable to escape.
When the gas inflow rate is limited and the outflow rate is zero, matter gradually accumulates in the BH feeding region owing to stellar mass loss.

The accumulation of matter leads to a higher density, and the gas eventually cools.
Cooling leads to a higher accretion rate, since the cooler gas does not counteract the gravity and since accretion is not inhibited by small-scale feedback, when the temperature is sub-virial.
The burst of accretion empties the feeding region. The accretion rate drops after the burst, and then a new phase of matter accumulation begins.
The accumulation phase of such accumulation-accretion cycles might reproduce the current state of NGC3115.
This possibility is to be explored with future time-dependent numerical simulations.

\subsubsection{Angular Momentum Transport}\label{subsubsec:angular_mom}
We did not explicitly treat angular momentum transport, which is partially justified a posteriori.
We find a relatively small circularization radius $r_{\rm circ}\lesssim0.05$~arcsec in the best-fitting solutions.
The accretion flow at $r=0.5-10$~arcsec probed with \textit{Chandra} might not feel the difference with an explicit treatment of the inner flow circularization.
The sonic point in the circularized flow is closer to the BH \citep{Popham:1998fp}, and the influence of conduction is expected to be stronger.
Then the shallow density profile is expected to continue down to several $r_{\rm g}$.
Having defined the injection of the angular momentum, we leave angular momentum transport and the inner flow connection for future work.

\subsection{Limitations of the Radial Solutions}\label{subsec:limitations_radial}
The presented modeling is performed under the strong approximation of one dimension.
The resultant treatment of gravitational forces is approximate and gas motions are restricted.

We compute the enclosed mass profile based on the surface brightness along the semi-major axis assuming zero ellipticity $\varepsilon=0$ of the stellar distribution.
However, \citet{Kormendy:1992ad} report an ellipticity of $\varepsilon\approx0.4$ at the Bondi radius. Since the ellipticity varies with radius
and the gravitational force is not trivially determined for a non-spherical mass distribution, we do not improve in this work upon the zero ellipticity approximation.

The gravitational force is generally lower in the case of non-zero ellipticity. To test the effect of a lower gravitational force
we search for a best-fitting solution with a smaller enclosed mass $M_{\rm enc,x}=0.7M_{\rm enc}$ and a fixed BH mass $1\times10^9M_\odot$.
We find the best-fitting advective solution with a higher normalization $f_q=0.59$ of the mass source function compared to $f_q=0.28$ for the $100\%$ of the enclosed mass.
The resultant normalization $f_q$ is much closer to unity, while $f_q$ varies little across the best-fitting solutions with different $M_{BH}$.
The effective supernova wind velocity is $v_{w,SN}=498{\rm km~s}^{-1}$ for the $70\%$ of the enclosed mass compared to $v_{w,SN}=455{\rm km~s}^{-1}$ for the $100\%$.
The correspondent change of the stagnation radius is from $r_{\rm st}=1.52$~arcsec to $r_{\rm st}=1.12$~arcsec.
The reduced chi-squared shows a small improvement by $\Delta(\chi^2/\dof)=4\times10^{-3}$.

The gas in the one-dimensional solution is restricted to either inflow or outflow radially.
More complex patterns may occur in two dimensions, such as inflow in the equatorial plane with outflow along the angular momentum axis.
While typical density and temperature profiles in two-dimensional solutions may be similar to those in one-dimensional solutions \citep{Yuan:2012lp,Sadowski:2013kc},
more detailed fitting of the data with the two-dimensional solutions is warranted. The lower gravitational force facilitates the outflow along the angular momentum axis.
This leads to an easier evacuation of the feeding region, so that the best-fitting two-dimensional solutions are expected to have a higher normalization $f_q$ of the mass loss rate.
Finding the self-consistent two-dimensional solutions might require the numerical simulations, and the present manuscript provides a starting point for such work.

\section{Acknowledgements}\label{sec:acknowledgements}
The authors thank Sam Leitner, Alexey Vikhlinin, Tassos Fragos, Sergey Nayakshin, Kazimierz Borkowski, Feng Yuan, Ranjan Vasudevan, and Richard Mushotzky for stimulating discussions
and the anonymous referee for useful suggestions. The work is supported by \textit{Chandra} XVP grant GO2-13104X. RVS is supported by NASA Hubble Fellowship grant HST-HF-51298.01.
RVS acknowledges hospitality of the Physics and Astronomy Department, University of North Carolina, Chapel Hill, where a part of the work was conducted.
\bibliographystyle{apj}
\bibliography{refs_fin}
\end{document}